\newcounter{myctr}

\documentclass{ws-ijqi}

\begin{document}

\markboth{H.-T. Elze}{Qubit exchange interactions from permutations of classical bits}


\title{QUBIT EXCHANGE INTERACTIONS FROM PERMUTATIONS OF CLASSICAL BITS}

\author{Hans-Thomas Elze}

\address{Dipartimento di Fisica ``Enrico Fermi'', Universit\`a di Pisa, \\ 
Largo Pontecorvo 3, I-56127 Pisa, Italia \\ 
elze@df.unipi.it}

\maketitle

\begin{history}
\received{31 October 2019}
\accepted{20 November 2019} 
\end{history}

\begin{abstract}
In order to prepare for the introduction of dynamical many-body and, eventually, field theoretical models, we show here that quantum mechanical exchange interactions in a three-spin chain can emerge from the deterministic dynamics of three classical Ising spins. States of the latter form an ontological basis, which will be discussed with reference to the ontology proposed in the   
{\it Cellular Automaton Interpretation of Quantum Mechanics} by 't\,Hooft. Our result illustrates a new Baker-Campbell-Hausdorff formula with terminating series expansion.   
\end{abstract}



\section{Introduction}  

In distinction to many studies of {\it quantum cellular automata} or {\it quantum walks}, be it for their interest in the foundations of (quantum) physics or in (information processing) applications, in this article we will {\it not} assume quantum mechanics as an ingredient from the beginning. Instead we will further discuss circumstances when quantum mechanical features can be found in the behaviour of certain kinds of classical cellular automata.  

We recall that ontological states have been proposed to underly quantum and, {\it a fortiori}, classical states of physical objects according to the {\it Cellular Automaton Interpretation of Quantum Mechanics} developed by  't\,Hooft.~\cite{tHooft2014}  

There is ample motivation to reexamine the foundations of quantum theory in the light of classical concepts -- concerning, above all, determinism and existence of ontological states of reality. Last not least, the Born rule and the infamous collapse of quantum states in measurement processes can find a surprising and intuitive explanation, if quantum states are regarded as mathematical objects. A different lucid argument supporting such a view has been given, {\it e.g.}, in a paper by Rovelli.~\cite{Rovelli2015} The quantum states represent mathematically fictitious superpositions of ontological (micro) states, while classical states are ontological (macro) states, or probabilistic superpositions thereof, as appropriate for nature's vast range of different scales.~\cite{tHooft2014}  

While the unification of General Relativity and Quantum Mechanics (QM) has not yet been achieved, it is a widely held belief that it will require drastic modifications of the respective foundations. The Cellular Automaton Interpretation (CAI) presents a comprehensive attempt to unveil a simpler structure beneath QM. 

We have recently studied so-called Hamiltonian cellular automata, comprising a large class of arguably very simple Cellular Automata 
(CA).~\cite{DMTimeMach12,PRA2014,EmQM13,Wigner13,Discrete14,Torino16} In particular, we have considered multipartite systems composed of simple two-state subsystems, akin to classical Ising models.~\cite{IJQI16}  They are candidates for ontological models underlying interacting many-body QM or quantum fields.~\cite{IJQI17} 

It must be emphasized that this goes beyond standard quantum theory by deforming structural elements in a specific way, such that textbook theory is recovered by taking a suitable continuum limit.~\cite{PRA2014} We refer to Refs.\,\cite{tHooft2014,PRA2014,IJQI17,FFM17} for further discussions, as well as to other related attempts.~\cite{H1,H2,H3,Elze,Kleinert,Groessing,Khrennikov,Margolus,Jizba,Mairi,Isidro,DArianoCA,Wetterich16a,Wetterich16b} 

It may be useful to distinguish here between {\it going beyond}, as indicated, and recent {\it reconstructions} from various alternative sets of axioms, without changing the contents of quantum theory, see, {\it e.g.}, Refs.\,\cite{DAriano,Hoehn1,Hoehn2,Hoehn3}. The latter are shedding light on QM by offering new options for experiments, for example, or by allowing to define and study precisely {\it generalized} (non-){\it quantum theories}. However, so far, they do not seem to address the possibility that quantum mechanics itself may be based on phenomena beneath that warrant the development of an encompassing theory founded on new principles.  

In the following Section\,2., we recapitulate the heuristic distinction between ontological, classical, and quantum states that we employ throughout this paper. 

We then present, in Section\,3., our study of the particular example of three 
Ising spins evolving linearly by permutations on their 3-bit space of eight states. 

This simple system fulfills the requirements of an ontological model. Yet we show that its dynamics can be described conveniently by a typical quantum mechanical Hamiltonian, incorporating Heisenberg exchange interactions. An important aspect of the underlying permutations is that they avoid the formation of would-be-quantum superposition states. Surprisingly, this aspect seems very well hidden when the 
{\it equivalent} QM language is used.    

We discuss these findings, which can be seen to follow from a new Baker-Campbell-Hausdorff formula, and indicate next steps in this program to build a {\it deterministic classical ontology for QM}. Last not least, it will be interesting to explore further how the violation of Bell's inequality can be understood in such detailed model cases;~\cite{Vervoort1,Vervoort2} that it is not a prohibitive issue in general has been argued by 't\,Hooft.~\cite{tHooft2014} 


\section{Matters of language -- distinguishing ontological, classical and quantum states}

To begin with, we emphasize that quantum states here are considered to form part of the mathematical language used and, thus, bear an epistemological character.~\cite{Rovelli2015} They are ``templates'' for the description of the reality beneath which, according to CAI, consists of ontological states following a deterministic rule of evolution.~\cite{tHooft2014}  

\vskip 0.1cm \noindent 
{\bf Definition 2.1.} \hskip 0.3cm
Ontological States  ($\cal OS$) 
\\ \phantom .\hskip 2.85cm 
are states that a closed physical system can be in. 
\vskip 0.1cm

The set of all such $\cal OS$ may be very large, possibly infinite. For simplicity, we assume that it is denumerable.

The physical reality ``out there'' comprises {\it no superpositions} of $\cal OS$. -- To construct a theory based on this attribute of $\cal OS$, which is essential, may seem wrong, in view of the overwhelming role played by superpositions in quantum theory. However, it must be stressed that quantum superpositions ``happen'' in the realm of the language used, {\it i.e.}, the mathematical formalism employed successfully to describe observed phenomena. 

Consequently, the $\cal OS$ can only evolve by {\it permutations} among themselves. Denoting $\cal OS$ by  $|A\rangle ,\; |B\rangle ,\; |C\rangle ,\; |D\rangle ,\;\dots\;$, for example, such a dynamics could be simply represented by:
\begin{equation} \label{PermEvol} 
|A\rangle\rightarrow |D\rangle\rightarrow |B\rangle\rightarrow\;\dots\;\;.
\end{equation} 
This kind of evolution is the only possible one, unless the set of states itself changes, {\it i.e.}, grows or shrinks.\,\footnote{While we do not consider a changing set of states here, this could be of interest when pondering the evolution of the Universe.}   

Formally, we may declare the $\cal OS$ to form a fixed orthonormal set, fixed once for all, and define a {\it Hilbert space} ${\cal H}$ with respect to this preferred basis.  -- Diagonal operators on this basis are {\it beables} and their eigenvalues characterize physical properties of the states, corresponding to the labels $A,\; B,\; C,\;\dots\;$ used above. 

The  association of the particular Hilbert space ${\cal H}$ with the set of $\cal OS$, then, leads to the following definition. 

\vskip 0.1cm \noindent 
{\bf Definition 2.2.} \hskip 0.3cm 
Quantum States ($\cal QS$) 
\\ \phantom .\hskip 2.85cm 
are superpositions of $\cal OS$, formally  introduced in ${\cal H}$. 
\vskip 0.1cm

These are  templates for doing  physics with the help of mathematics. --  
The amplitudes that specify a $\cal QS$ need to be interpreted, when describing experiments in terms of these states. By experience, interpreting amplitudes in terms of probabilities has been an extraordinarily useful invention. Thus, the {\it Born rule} is built in by definition!\,\footnote{Furthermore, the Born rule can be related to a counting procedure and conserved two-time function of CA, which generalizes the norm of $\cal QS$.\,\cite{FFM17}} One might consider instead to abandon the proportionality  between {\it absolute values squared} of complex amplitudes and probabilities, which is not forbidden by any element of quantum theory, however, would unnecessarily complicate the available mathematical tools.~\cite{tHooft2014}     

We anticipate that generally it will not be easy to relate unitary evolution of $\cal OS$ by permutations, as in (\ref{PermEvol}), to a more or less familiar looking Hamilton operator, in particular in the presence of interactions.~\cite{tHooft2014,IJQI16,IJQI17,FFM17,Wetterich16a,Wetterich16b} Which motivates the study of the model of Section\,3.    

In any case, we recognize a good part of the machinery of quantum theory already in place, including the powerful means of unitary transformations in Hilbert spaces, but with the new perspective furnished by the existence of $\cal OS$ of reality. 

Finally, we define {\it classical states} in relation to $\cal OS$. -- Usually, they are thought to describe certain limiting situations of QM, especially in the presence of environment induced decoherence, {\it i.e.}, when the object under study is part of a larger interacting system. They have formed the realm of classical physics before the advent of QM. -- Within the CAI, however, classical states belong to deterministic macroscopic systems, including billiard balls, pointers of apparatus, planets, etc., {\it and} are formed of ontological states that are not resolved individually. 

\vskip 0.1cm \noindent 
{\bf Definition 2.3.} \hskip 0.3cm 
Classical States ($\cal CS$) 
\\ \phantom .\hskip 2.85cm 
of a closed physical (macro)system are  
\\ \phantom .\hskip 2.85cm 
probabilistic distributions of its $\cal OS$. 
\vskip 0.1cm 

Repeatedly performed experiments or any kind of {\it approximately} repeating evolution of a sufficiently but never completely isolated component of the overall system, say, the Universe,  
pick up different initial conditions regarding the $\cal OS$. 
Therefore, a suitable {\it classical apparatus} forming part of such situations must generally be expected to yield different pointer positions as outcomes. 

Furthermore, then, the  
probability of a particular outcome directly reflects the probability of having a particular $\cal OS$ as initial condition. This conservation law, the {\it Conservation of Ontology}, follows directly from the absence of 
superpositions ``out there'' and the evolution of 
$\cal OS$ by permutations among themselves.~\cite{tHooft2014} Since, using  quantum superpositions of $\cal OS$ to describe the initial state approximately, as good as possible, we obtain for an evolving $\cal QS$ 
$|Q\rangle$: 
\begin{eqnarray} \label{QS} 
|Q\rangle &:=&\alpha |A\rangle +\delta |D\rangle +\dots  
\;\;,\;|\alpha |^2+|\delta |^2 +\dots\;  
=1\;\;, 
\\ [1ex] \label{QSEvol}
\mbox{then},&\phantom .& 
|Q\rangle\;\longrightarrow\;
\alpha |D\rangle +\delta |B\rangle +\dots  
\;\;, \end{eqnarray} 
{\it i.e.}, the amplitudes remain the initial ones, while the $\cal OS$ evolve by  permutations, {\it e.g.} as in  (\ref{PermEvol}). 

Hence, the reduction or collapse to a $\delta$-peaked distribution of pointer positions, the core of the {\it measurement problem}, is an apparent effect. It arises due to the intermediary use of {\it quantum mechanical templates}, in particular superposition states such as $|Q\rangle$, when describing the evolution of what in reality are $\cal OS$ that differ in different runs of an experiment, either $|A\rangle$ or $|D\rangle$ or $\dots$ in the example of (\ref{QS}). 

According to CAI, superpositions of $\cal OS$ do not exist 
``out there'' and, therefore, {\it no collapse} or reduction to one of their components can occur!\,\cite{tHooft2014}  

This does, of course, not imply that quantum mechanical superposition states are to be avoided. On the contrary, part of the motivation for CAI and its perspective is to better understand, why they are so extremely effective in describing experiments probing nature. 

It is encouraging for attempts to formulate an ontological theory that QM {\it per se} does not need any stochastic or nonlinear reduction process, which could modify the collapse-free linear unitary evolution. Measurements, 
according to CAI, are simply {\it interactions} between the degrees of freedom belonging to an object and those belonging to an apparatus, altogether evolving through ontological states. While the elegance and simplicity of this point of view can hardly be denied, the construction of examples of interacting systems is not entirely  straightforward.  


\section{Permutations of classical bits with a QM Hamiltonian} 

We considered a discrete dynamical theory before that deviates drastically from quantum theory, at first sight. With the help of Sampling Theory,\,\cite{Shannon,Jerri} however, it has been shown that members of this class of {\it Hamiltonian CA} are mapped one-to-one to continuum models of nonrelativistic QM, in which a deformation through a new time or length scale  enters.\,\cite{PRA2014,EmQM13,Discrete14,TDLee}    

Yet all these models have been one-body models, {\it i.e.} with forces that are not dynamical but external, described by an external potential in the Hamiltonian. This has led to consider 
{\it multipartite systems} in this context, aiming to arrive at {\it interacting} discrete many-body or field models.\,\cite{IJQI16,IJQI17}   
 
Presently, we continue in this line, however, refer specifically to an ontological model with $\cal OS$ that evolve by permutations, as discussed in Section\,2.  


\subsection{Some properties of permutations} 

Let $N$ objects, $A_1,A_2,\dots ,A_N$ (``states''), be mapped in $N$ steps onto one another, involving {\it all} states exactly once. 
This can be represented by {\it unitary} $N\times N$ matrices with {\it one} off-diagonal arbitrary phase per column and row and zero elsewhere. 

For example, consider: 
\begin{equation} \label{U3} 
\hat U_3:=\left (\begin{array}{c c c} 
0 & e^{-i\phi_1} & 0 \\ 
0 & 0 & e^{-i\phi_2} \\ 
e^{-i\phi_3} & 0 & 0 \\ 
\end{array}\right )
\;\;,\;\;\; 
\hat U_3\hat U_3^\dagger =\mathbf{1} 
\;\;. \end{equation}   
This permutation matrix has the property: 
\begin{equation} \label{U33} 
(\hat U_3)^3=
e^{-i(\phi_1+\phi_2+\phi_3)}\;\mathbf{1}
\;\;, \end{equation} 
which immediately yields its eigenvalues as three 
roots of 1, which lie on the unit circle in the complex plane, multiplied by an overall phase. 	
	
Similarly, one finds for unitary $N\times N$ matrices that represent permutations: 
\begin{equation} \label{UNN}  
(\hat U_N)^N=
\exp (-i\sum_{k=1}^N\phi_k)\;\mathbf{1}
\;\;. \end{equation} 
Defining the {\it Hamiltonian} by: 
\begin{equation} \label{Hop} 
\hat U_N=:e^{-i\hat H_NT} 
\;\;, \end{equation} 
with $T$ introducing a time scale, 
its eigenvalues follow from Eq.\,(\ref{UNN}). This  
results in the diagonal form 
of the Hamiltonian:   
\begin{equation} \label{Hdiag} 
\hat H_{N,diag}=\mbox{diag}\left (\frac{1}{NT}(
\sum_{k=1}^N\phi_k+2\pi n)\;|\;n=0,\;\dots\; ,N-1
\right ) 
\;\;. \end{equation} 
  
In passing we mention that these simple results 
form the basis of so-called 
{\it cogwheel models}.~\cite{tHooft2014} 
These deterministic models show interesting quantum mechanical features, 
when considered in suitable limiting situations.~\cite{Elze,ElzeRelativPart} 
For $N\rightarrow\infty$ and $T\rightarrow 0$, 
with $NT\equiv\omega^{-1}$ fixed,  
a {\it quantum harmonic oscillator} is obtained,  
while $\omega\rightarrow 0$ yields a  
{\it free quantum particle}. 

As mentioned, difficulties are encountered, when 
one tries to introduce interactions among such 
simplest building blocks, which by themselves end up 
to represent quantum mechanical one-body systems. 


\subsection{Ising spins as ontological degrees of freedom} 

Somehow, besides ``global'' evolution of $\cal OS$ by permutations, we need to have some 
additional ``local'' or internal structure of the states, in 
order to construct more interesting dynamical models.  

Classical {\it two-state Ising spins} (Boolean variables or bits) are suitable elementary objects to compose a multipartite CA. For simplicity, we consider only a chain of  
three coupled Ising spins ``1,2,3'', at present. 

Interactions among the three spins, leading to permutations among their $2^3$ possible $\cal OS$, will be generated 
by {\it spin exchange}, a permutation 
$\hat P_{ij}\;(\equiv\hat P_{ji})$ involving two spins, labeled $i,j=1,2,3$, with the following properties: 
\begin{equation} \label{Pijdef} 
\hat P_{ij}|s_i,s_j\rangle :=
|s_j,s_i\rangle 
\;\;,\;\;\; \hat P_{ji}\hat P_{ij}
=(\hat P_{ij})^2=\mathbf{1}
\;\;, \end{equation} 
where the states of a single spin are 
$s_k=\pm 1$ or, graphically, $s_k=\uparrow ,\downarrow$, for ``spin up, spin down'', respectively; we use the ket notation $|s_i,s_j\rangle$ 
to indicate that the first spin has value $s_i$, the second value $s_j$, and similarly for all three spins. 

If we identify the two states $s_k=\pm 1$ of an Ising spin with the eigenstates of the 
Pauli matrix $\hat\sigma^z$, 
$\psi_+=(1,0)^t$ and $\psi_-(0,1)^t$, respectively, then the unitary operator 
$\hat P_{ij}$ can be expressed in terms 
of the Pauli spin-1/2 matrices: 
\begin{equation} \label{PijPauli}
\hat P_{ij}=\frac{1}{2}(\underline{\hat\sigma}_i
\cdot\underline{\hat\sigma}_j+\mathbf{1})
\;\;, \end{equation} 
where $\underline{\hat\sigma}$ denotes 
the vector formed by $\hat\sigma^x,\hat\sigma^y,\hat\sigma^z$. 
This hints at a relation with QM of qubits, to which we shall come back in the following.

Finally, an elementary example suffices to show: 
\begin{equation} \label{comm} 
[\hat P_{ij},\hat P_{jk}]\neq 0\;\;, \;\;
\mbox{for}\;\; i\neq k
\;\;, \end{equation} 
where no summation over $j$ is implied.   


\subsection{Dynamics} 
 
Our next task is to define the unitary  operator $\hat U$ that evolves the state of the 
three classical spins under consideration 
in a discrete time step $T$ 
and extract the corresponding Hamiltonian $\hat H$, 
{\it cf.} Eqs.\,(\ref{Hop})-(\ref{Hdiag}).  
A suitable choice may simply be: 
\begin{equation} \label{U} 
\hat U:=\hat P_{12}\hat P_{23}=:\exp (-i\hat HT) 
\;\;, \end{equation} 
acting sequentially on the indicated pairs 
of spins. 

An important simplification arises, because the {\it numbers of up and down spins are conserved} when the interaction described by 
Eq.\,(\ref{U}) acts on one of the $\cal OS$. Therefore, we expect 
$\hat U$ to have a block diagonal 
structure. 

Let us order the eight $\cal OS$ in the following way: 
$|1\rangle :=|\uparrow ,\uparrow ,\uparrow\rangle$, 
$|2\rangle :=|\uparrow ,\uparrow ,\downarrow\rangle$, 
$|3\rangle :=|\uparrow ,\downarrow ,\uparrow\rangle$, 
$|4\rangle :=|\downarrow ,\uparrow ,\uparrow\rangle$, 
$|5\rangle :=|\downarrow ,\downarrow ,\uparrow\rangle$, 
$|6\rangle :=|\downarrow ,\uparrow ,\downarrow\rangle$,
$|7\rangle :=|\uparrow ,\downarrow ,\downarrow\rangle$,  
$|8\rangle :=|\downarrow ,\downarrow ,\downarrow\rangle$. Then, indeed, the update operator $\hat U$ of Eq.\,(\ref{U}) can be represented by the following block diagonal 
$8\times 8$ matrix: 
\begin{equation} \label{Ublock}  
\hat U=\left (\begin{array}{c c c c} 
1 & \phantom . & \phantom . & \phantom . \\ 
\phantom . & \hat U_3 & \phantom . & \phantom . \\ 
\phantom . & \phantom . & \hat U_3 & \phantom . \\ 
\phantom . & \phantom . & \phantom . & 1 \\ 
\end{array}\right ) 
\;\;, \end{equation} 
where $\hat U_3$ is the unitary $3\times 3$ matrix defined in (\ref{U3}), with $\phi_1=\phi_2=\phi_3=0$ henceforth; all 
other matrix elements not explicitly given are zero.  

Similarly as before, {\it cf.} Eq.\,(\ref{Hdiag}), since $(\hat U)^3=\mathbf{1}$, we immediately obtain also the diagonal form of the  
Hamiltonian defined in Eqs.\,(\ref{U}): 
\begin{equation} \label{Hdiagonal} 
\hat H_{diag}=\frac{2\pi}{3T}\cdot 
\mbox{diag}\left (0,\;0,1,2,\;0,1,2,\;0
\right )  
\;\;. \end{equation} 
We note the degeneracy of the eigenvalues.  

Furthermore, due to the simple structure of $\hat U$, it is straightforward to find its eigenstates. The eigenstates corresponding to two of the zero eigenvalues of $\hat H_{diag}$ are simply  the states $|1\rangle$ and $|8\rangle$ defined before Eq.\,(\ref{Ublock}), with all spins either up or down. -- Next, we construct the eigenstates of $\hat U_3$. It is convenient to introduce an auxiliary basis, $|\alpha\rangle :=(1,0,0)^t$, $|\beta\rangle :=(0,1,0)^t$, $|\gamma\rangle :=(0,0,1)^t$. In terms of these orthonormal vectors, one finds the normalized eigenvectors: 
\begin{eqnarray} \label{eigen}  
|A\rangle &=&(|\alpha\rangle +|\beta\rangle +|\gamma\rangle )/\sqrt 3 
\;\;, \nonumber \\ [1ex]  
|B\rangle &=&(|\alpha\rangle +e^{-2\pi i/3}|\beta\rangle +e^{2\pi i/3}|\gamma\rangle )/\sqrt 3  
\;\;, \nonumber \\ [1ex]  
|C\rangle &=&(|\alpha\rangle +e^{2\pi i/3}|\beta\rangle +e^{-2\pi i/3}|\gamma\rangle )/\sqrt 3  
\;\;, \end{eqnarray} 
which obey $\hat U_3|A\rangle =|A\rangle$, 
$\hat U_3|B\rangle =e^{-2\pi i/3}|B\rangle$, 
$\hat U_3|C\rangle =e^{-4\pi i/3}|C\rangle$, 
with the correct eigenvalues of $\hat U_3$. 
The unitary diagonalizing matrix $D$, which maps 
$\{ |\alpha\rangle ,|\beta\rangle, |\gamma\rangle\}$ to $\{ |A\rangle ,|B\rangle ,|C\rangle\}$, 
is given by: 
\begin{equation} \label{D} 
\hat D=\frac{1}{\sqrt 3}
\left (\begin{array}{c c c} 
1 & 1 & 1 \\ 
1 & e^{-2\pi i/3} & e^{2\pi i/3} \\ 
1 & e^{2\pi i/3} & e^{-2\pi i/3} \\ 
\end{array}\right )
\;\;. \end{equation} 
It serves us to map part of the diagonalized Hamiltonian, which corresponds to one of the $\hat U_3$ blocks, namely  
$\hat H_{3,diag}:=\mbox{diag}\left (0,1,2\right )\cdot\frac{2\pi}{3T}$, back to the corresponding part 
$\hat H_{aux}$ of $\hat H$ which acts on the auxiliary basis 
$\{ |\alpha\rangle ,|\beta\rangle, |\gamma\rangle\}$: 
\begin{eqnarray} \label{Haux} 
\hat H_{aux}&=&\hat D^\dagger\hat H_{3,\;diag}\hat D
\nonumber \\ [1ex] 
&=&\frac{2\pi}{3T}
\left (\begin{array}{c c c} 
1 & c & c^* \\ 
c^* & 1 & c \\ 
c & c^* & 1 \\ 
\end{array}\right ) 
\;\;,\;\;\;\mbox{with}\;\; c:=-\frac{1}{2}
+\frac{i}{2\sqrt 3}
\;\;. \end{eqnarray}
Note the permutation of entries between one   
row or column and the next. This Hamiltonian, of course, appeared also 
in the analysis of the $N=3$ cogwheel model.~\cite{tHooft2014}  

The crucial step, in order to arrive at the Hamiltonian 
$\hat H$ of Eq.\,(\ref{U}), is to identify the auxiliary states with corresponding $\cal OS$, which were listed 
before Eq.\,(\ref{Ublock}), and to express 
$\hat H_{aux}$ in terms of operators acting on the three Ising spins ``1,2,3'', of which the ontological states 
are composed. This works as follows.
  
Let us identify $|\alpha\rangle\equiv |\uparrow ,\uparrow , \downarrow\rangle$, $|\beta\rangle\equiv |\uparrow ,\downarrow , \uparrow\rangle$, and $|\gamma\rangle\equiv |\downarrow ,\uparrow , \uparrow\rangle$; we recall that in the last case, {\it e.g.}, the notation means spin ``1'' down, spins ``2'' and ``3'' up. Thus, with Eq.\,(\ref{Haux}) the following identification is obtained, for example: 
\begin{eqnarray} \label{Hauxcorr} 
\hat H_{aux}|\beta\rangle &=&\frac{2\pi}{3T}\big (c|\alpha\rangle +|\beta\rangle +c^*|\gamma\rangle \big )  
\nonumber \\ [1ex] 
&\equiv&\frac{2\pi}{3T}\big (c\hat P_{23}+\mathbf{1}+c^*\hat P_{12} \big )|\uparrow ,\downarrow , \uparrow\rangle 
\;\;, \end{eqnarray}
using the spin exchange operators introduced in Eq.\,(\ref{Pijdef}); similar  
identifications follow for the other two members of the auxiliary basis, 
\begin{eqnarray} \label{Hauxcorr1}  
\hat H_{aux}|\alpha\rangle 
&\equiv&\frac{2\pi}{3T}\big (\mathbf{1}+c^*\hat P_{23}+c\hat P_{13} \big )|\uparrow ,\uparrow ,\downarrow\rangle 
\;\;, 
\\ [1ex] \label{Hauxcorr2} 
\hat H_{aux}|\gamma\rangle 
&\equiv&\frac{2\pi}{3T}\big (c^*\hat P_{13}+c\hat P_{12}+\mathbf{1} \big )|\downarrow ,\uparrow ,\uparrow\rangle 
\;\;, \end{eqnarray} 
and, correspondingly, for an  
auxiliary Hamiltonian related to the second 
$3\times3$ block entering $\hat U$ in Eq.\,(\ref{Ublock}). 

Finally, we recall that the rows or columns of 
$\hat H_{aux}$ are simply related by cyclic permutations. For a generic $\cal OS$, say 
$|x,y,z\rangle$, with $x,y,z=\uparrow,\downarrow$, such permutations can be represented by: 
\begin{equation} \label{permutations} 
\hat P_{13}\hat P_{23}|x,y,z\rangle =|y,z,x\rangle 
\;\;,\;\; 
(\hat P_{13}\hat P_{23})^2|x,y,z\rangle =|z,x,y\rangle 
\;\;,\;\; 
\end{equation} 
and $(\hat P_{13}\hat P_{23})^3=\mathbf{1}$. 
This allows to write the ($8\times 8$ matrix) Hamiltonian $\hat H$ of Eq.\,(\ref{U}), which incorporates especially  Eqs.\,(\ref{Hauxcorr})-(\ref{Hauxcorr2}) but acts on the eight $\cal OS$, in a concise form: 
\begin{eqnarray} \label{Hresult} 
\hat H&=&\frac{2\pi}{3T}\big ( 
\mathbf{1}+c^*\hat P_{13}\hat P_{23}+c(\hat P_{13}\hat P_{23})^2\big ) 
\nonumber \\ [1ex] 
&=&\frac{2\pi}{3T}\big ( 
\mathbf{1}+c\hat P_{23}\hat P_{13}+c^*\hat P_{13}\hat P_{23}\big )
\;\;, \end{eqnarray} 
which is evidently self-adjoint; note that 
$(\hat P_{13}\hat P_{23})^\dagger 
=\hat P_{23}\hat P_{13}=(\hat P_{13}\hat P_{23})^2$. Furthermore, it is noteworthy  that all parts of $\hat H$ commute with each other, since $[\hat P_{23}\hat P_{13},\hat P_{13}\hat P_{23}]=\mathbf{1}-\mathbf{1}$ in particular. 

The explicit form of the Hamiltonian, Eq.\,(\ref{Hresult}), presents our main result. In the following, we will discuss some of its implications. 


\subsection{A Baker-Campbell-Hausdorff formula with a qubit Hamiltonian for bits} 

Here, we first look at an interesting formal aspect of our result describing the dynamics by permutations of $\cal OS$ by a Hamilton operator. Returning to Eq.\,(\ref{U}), together with Eq.\,(\ref{Pijdef}), we see that the two permutations composing the evolution operator $\hat U$ can be 
exponentiated separately as follows: 
\begin{equation} \label{exp} 
\hat P_{ij}=i\exp (-i\frac{\pi}{2}\hat P_{ij}) 
\;\;, \end{equation} 
using $(\hat P_{ij})^2=\mathbf{1}$. However, since $[\hat P_{12},\hat P_{23}]\neq 0$, we cannot evaluate 
the Hamiltonian $\hat H$ by simply  adding the exponents 
obtained with the help of Eq.\,(\ref{exp}). To put it differently, we would like to know, what is 
$$-i\hat HT=\log (\hat P_{12}\hat P_{23}) 
\;\;, $$ 
for two noncommuting permutations acting on three Ising spins (or bits). 

This kind of algebraic problem with noncommuting operators is familiar from QM or Lie group theory.   
Let $\exp (X)\exp (Y)=\exp (Z)$. Then, a  formal solution for $Z$ in terms of $X,Y$ is provided by the {\it Baker-Campbell-Hausdorff formula} (BCH):   
\begin{equation} \label{BCH} 
Z=X+Y+\frac{1}{2}[X,Y]+\frac{1}{12}
([X,[X,Y]]+[Y,[Y,X]])  
-\frac{1}{24}[Y,[X,[X,Y]]]
+\;\dots
\;\;, \end{equation}  
{\it i.e.}, a series expansion in terms 
of increasingly complicated iterated commutators. The coefficients of the series are known, yet it is generally difficult to ascertain its convergence. Several exceptional cases are known when the series terminates. In recent works by Visser {\it et al.} and by Matone, 
with references to earlier work, some interesting new classes of such finite solutions have been constructed.~\cite{Visser,Matone}   
 
Instead, in the preceding section, rather elementary considerations have lead us to results which can be summarized by the new terminating BCH formula:  
\begin{equation} \label{BCHnew} 
i^2\exp\big (-i\frac{\pi}{2}\hat P_{12}\big )  
\exp\big (-i\frac{\pi}{2}\hat P_{23}\big )= 
\exp\big (-i\frac{2\pi}{3}(\mathbf{1}+c\hat P_{23}\hat P_{13}+c^*\hat P_{13}\hat P_{23})\big )  
\;\;, \end{equation} 	
using Eqs.\,(\ref{U}), (\ref{Hresult}),  (\ref{exp}), and with $c$ from (\ref{Haux}); the coefficients of $\pi /2$ in this formula can be modified by adding integer multiples of $2\pi$, without changing the result. --
To obtain this result from the general BCH formula (\ref{BCH}), does not seem impossible but rather complicated.  

In passing we mention that the right-hand side of Eq.\,(\ref{BCHnew}) can be factorized: 
\begin{equation} \label{BCHnew1} 
\dots\;=\;
\exp \big (-i\frac{2\pi}{3}\big ) 
\exp \big (-i\frac{2\pi}{3}c\hat P_{23}\hat P_{13}\big )
\exp \big (-i\frac{2\pi}{3}c^*\hat P_{13}\hat P_{23}\big )  
\;\;, \end{equation}  
since all terms in the exponent commute. 

It will be interesting to find out, how our derivation can be generalized, {\it e.g.}, for more than three Ising spins (or bits). 

Next, we recall the relation between  exchange operations (permutations) and Pauli matrices, Eq.\,(\ref{PijPauli}), 
$\hat P_{ij}=(\underline{\hat\sigma}_i\cdot\underline{\hat\sigma}_j+\mathbf{1})/2$. If we now allow the Ising spins, on which the 
Hamiltonian $\hat H$ of Eq.\,(\ref{Hresult}) acts, to be embedded into the larger Hilbert space of three qubits, then $\hat H$ expressed in terms of the appropriate Pauli matrices can be considered as a genuine {\it quantum mechanical operator}.  

This brings about an interesting situation: \\ \noindent  
Equations like (\ref{U}) together with 
(\ref{Hresult}), as well as Eq.\,(\ref{BCH}), are only valid with their precisely determined numerical coefficients. Furthermore, despite the quantum mechanical appearance of the Hamiltonian, the operator $\hat U$ describes the {\it evolution of ontological states}. In particular, {\it no superpositions} of $\cal OS$ are produced, as it should be according to the discussion in Section\,2. 

Then, let us envision a realistic model of some sufficiently isolated part of the Universe which works along these lines -- and someone in search of such an ontological model. By necessarily limited experimental means this physicist will not be able to determine all relevant dimensionless (coupling or charge) constants precisely.\,\footnote{Symmetry arguments should help. While still little is known or can justifiably be assumed about symmetry principles that operate at the level of $\cal OS$, 't\,Hooft's discussion of ``Demands and Rules'' that an ontological model of the Universe should obey presents  steps in this direction.\cite{tHooftScard}} 

In analogy to what would happen in the case of our present toy model, a resulting approximate Hamiltonian, except if it is diagonal and rather uninteresting, will produce unphysical superpositions of ontological states!   

This can be illustrated simply, {\it e.g.}, by perturbing the right-hand side of Eq.\,(\ref{exp}): 
$$ 
i\exp (-i\frac{\pi}{2}(1+\epsilon )\hat P_{ij}) 
=\hat P_{ij}-i\frac{\pi}{2}\epsilon\cdot\mathbf{1}
+\mbox{O}(\epsilon^2)\;\;,\;\;
0<\epsilon\ll 1\;\;.$$  The resulting sum of terms unavoidably creates superposition states, which lie in a Hilbert space for qubits rather than concerning Ising spins (or bits). Similarly, any approximation to the Hamiltonian of Eq.\,(\ref{Hresult}) or perturbation of the BCH formula (\ref{BCHnew}) likely will produce apparent QM effects.   

We conclude that an only {\it approximately known  ontological Hamiltonian must lead to   misinterpretation}, namely that the system under study 
behaves {\it quantum mechanically},  
due to the presence of superpositions of $\cal OS$.

A surprise worth further exploration. 
 

\section{Conclusions} 

We discuss the notion of {\it Ontological States} ($\cal OS$) in the context of a composite system consisting of three classical {\it Ising spins} (or bits) and its dynamics. The $\cal OS$ have been introduced as the  basis on which models possibly underlying quantum mechanical ones must be  built.\,\cite{tHooft2014} 

Characteristic for $\cal OS$ -- the states a physical system can be in -- is that they evolve deterministically by {\it permutations} among themselves, since the familiar superposition states appearing in QM belong to the mathematical theory describing experimental findings, but are not considered to exist ``out there''.     

Single-component Cellular Automata that we have studied earlier allowed to 
reconstruct the dynamics of one-body models in QM with external potentials in terms of 
deterministic ones that are characterized by a finite discreteness scale. 

One is led to ask, whether in this setting, generally, there is room for $\cal OS$ and their particular permutation dynamics when it comes to interacting 
multipartite systems, {\it i.e.} many-body systems or fields. 

To progress towards such more interesting complex situations, we propose to begin with systems composed of two-state subsystems, such as Ising spins. In particular, we presently study a three-spin chain and its discrete dynamics generated by permutations among the $2^3=8$ possible $\cal OS$. 

We show that this opens new possibilities, since the unitary evolution of the system can be described by  a Hamilton operator, which appears to be of quantum mechanical kind. It incorporates spin exchange 
interactions in a nontrivial way.\,\footnote{They are known, {\it e.g.}, in the Heisenberg model of ferromagnetism.}    

Nevertheless, this evolution law 
with the derived exact Hamiltonian does not produce 
superpositions of $\cal OS$. This is implied 
by a new {\it Baker-Campbell-Hausdorff formula} 
with finitely many terms that we obtain. It describes the underlying permutations by 
the exponential of an equivalent Hamiltonian. 
  
We discuss that any approximation of the 
Hamiltonian, an inaccuracy of the fixed coupling constants in particular, would unavoidably lead from $\cal OS$ to 
superposition states and, thus, generate, 
typical quantum mechanical behaviour, of qubits in the present case. Which naturally opens the way for some speculations. 

It should be interesting to extend our study to  multipartite systems, {\it e.g.}, by dividing a given number of Ising 
spins in two sets and ask under what conditions, or rather approximations, one can observe entanglement between subsystems. We emphasize that presently the discrete configuration space of ontological degrees of freedom (Ising spins or bits) is strictly smaller than the corresponding continuous Hilbert space of qubits. As we discussed, the latter is opened up `by mistake' -- namely, when the effects of QM arise due to an inaccurate treatment of the ontological dynamics generated by permutations. This means, by approximating the exact BCH formula. We leave this important question for future work. 

However, to distinguish the role of ontic {\it vs.} epistemic (QM) aspects in complex situations can be very relevant, for example, when 
the formalism of quantum theory is successfully applied to situations that traditionally fall outside of physics.~\cite{AtmanspacherPrimas,Khrennikov2} It may sometimes even be fruitful to view ontic and epistemic features of a system side by side, such as represented by the conformational and functional aspects 
of complex molecules.~\cite{Khrennikov1} Furthermore, the longstanding question whether classical (ontological, {\it cf.} Section\,2.) and quantum mechanical degrees of freedom can consistently be combined in one dynamical scheme deserves reconsideration in the light of the present discussion, 
see Ref.~\cite{Elzehybrid} with references to earlier work.   

Next steps to be considered should include   
the extension to larger systems, in order to see whether the present results can be generalized in a straightforward way, with respect to dimensionality, but also incorporating the finite maximal signal velocity demanded by special relativity. Furthermore, the prerequisites for models with internal symmetries need to be understood.  

\section*{Acknowledgments}
It is a pleasure to thank A. Elitzur, 
G. 't\,Hooft, and L. Vervoort for discussions, 
M. Genovese for kind invitation to 
{\it Quantum 2019} (Torino, May 25 - June 1, 2019), 
and C. Wetterich for discussions and kind hospitality at Institut f\" ur Theoretische Physik (Heidelberg). Thanks are due to a referee for his interesting questions and  helpful remarks concerning some references.  





\begin{thebibliography}{99}

\bibitem{tHooft2014} G. 't Hooft, The Cellular Automaton Interpretation of Quantum Mechanics. 
{\em Fundamental Theories of Physics} {\bf 185} (Springer International Publishing, 2016). 

\bibitem{Rovelli2015} C. Rovelli, An argument against the realistic interpretation of the wave function. {\it Preprint} arXiv:1508.05533v2 [quant-ph] (2015). 

\bibitem{DMTimeMach12} H.-T. Elze, Discrete mechanics, time machines and hybrid systems.  
{\em EPJ Web of Conferences} {\bf 58} (2013) 01013.  

\bibitem{PRA2014} H.-T. Elze, Action principle for cellular automata and the linearity of quantum mechanics.  
{\em Phys. Rev. A} {\bf 89} (2014) 012111.       

\bibitem{EmQM13} H.-T. Elze, The linearity of quantum mechanics from the perspective of 
Hamiltonian cellular automata.  
{\em J. Phys.: Conf. Ser.} {\bf 504} (2014) 012004. 

\bibitem{Wigner13} H.-T. Elze, Quantumness of discrete Hamiltonian cellular automata.  
{\em EPJ Web of Conferences} {\bf 78} (2014) 02005.  

\bibitem{Discrete14} H.-T. Elze, Are nonlinear discrete cellular automata compatible with quantum mechanics? 
{\em J. Phys.: Conf. Ser.} {\bf 631} (2015) 012069.  

\bibitem{Torino16} H.-T. Elze, Quantum models as classical cellular automata.  
{\em J. Phys.: Conf. Ser.} {\bf 845} (2017) 012022.  

\bibitem{IJQI16} H.-T. Elze, Multipartite cellular automata and the superposition principle.  
{\em Int. J. Quant. Info. (IJQI)} {\bf 14} (2016) 1640001. 

\bibitem{IJQI17} H.-T. Elze, Ontological states and dynamics of discrete (pre-) quantum systems. {\em Int. J. Quant. Info. (IJQI)} {\bf 15} (2017) 1740013. 

\bibitem{FFM17} H.-T. Elze, On configuration space, Born's rule and ontological states. Invited contribution to the {\em FIAS Interdisciplinary Science Series} volume dedicated to Walter Greiner, to appear; {\it preprint} arXiv:1802.07189 [quant-ph] (2018).  

%

\bibitem{H1} G. 't Hooft, Quantization of discrete deterministic theories by Hilbert space extension,
{\em Nucl. Phys. B} {\bf 342} (1990) 471.  

\bibitem{H2} G. 't Hooft, K. Isler and S. Kalitzin, Quantum field theoretic behavior of a deterministic cellular automaton. 
{\em Nucl. Phys. B} {\bf 386} (1992) 495. 

\bibitem{H3} G. 't Hooft, Quantummechanical behaviour in a deterministic model.  
{\em Found. Phys. Lett.} {\bf 10} (1997) 105. 

\bibitem{Elze} H.-T. Elze and O. Schipper,   
Time without time: a stochastic clock model.  
{\em Phys. Rev. D} {\bf 66} (2002) 044020.  

\bibitem{Kleinert} Z. Haba and H. Kleinert,   
Towards a simulation of quantum computers by classical systems.  
{\em Phys. Lett. A} {\bf 294} (2002) 139. 

\bibitem{Groessing} G. Gr\"ossing, From classical Hamiltonian flow to quantum theory:  derivation of the Schr\"odinger equation.  
{\em Found. Phys. Lett.} {\bf 17} (2004) 343.  

\bibitem{Khrennikov} A. Khrennikov, Generalizations of quantum mechanics induced by classical statistical field theory.  
{\em Found. Phys. Lett.} {\bf 18} (2005) 637. 

\bibitem{Margolus}  N. Margolus,  
Mechanical systems that are both classical and quantum. 
{\em Lect. Notes Comput. Sc.} {\bf 9099} (2015) 169; 
see also {\it preprint} arXiv:0805.3357v2 [quant-phys] (2008). 

\bibitem{Jizba} M. Blasone, P. Jizba, F. Scardigli and G. Vitiello,   
Dissipation and quantization for composite systems.  
{\em Phys. Lett. A} {\bf 373} (2009) 4106.  

\bibitem{Mairi} M. Sakellariadou, A. Stabile and G. Vitiello,   
Noncommutative spectral geometry, algebra doubling and the seeds of quantization.  
{\em Phys. Rev. D} {\bf 84} (2011) 045026. 

\bibitem{Isidro} D. Acosta, P. Fernandez de Cordoba, J.M. Isidro and J.L.G. Santander,  
An entropic picture of emergent quantum mechanics.  
{\em Int. J. Geom. Meth. Mod. Phys.} {\bf 9} (2012) 1250048. 

\bibitem{DArianoCA} A. Bisio, G.M. D'Ariano and A. Tosini, Quantum field as a quantum cellular automaton: the Dirac free  evolution in one dimension.  
{\em Ann. Phys.} {\bf 354} (2015) 244. 

\bibitem{Wetterich16a} C. Wetterich, Information transport in classical statistical systems.  
{\it Preprint} arXiv:1611.04820v4  [cond-mat.stat-mech] (2016). 

\bibitem{Wetterich16b} C. Wetterich, Fermions as generalized Ising models.  
{\it Preprint} arXiv:1612.06695v2 [cond-mat.stat-mech] (2016).  

%

\bibitem{DAriano} G. Chiribella, G.M. D'Ariano and P. Perinotti, Informational derivation of Quantum Theory. 
{\em Phys. Rev. A} {\bf 84} (2011) 012311.  

\bibitem{Hoehn1}  P.A. Hoehn, Reflections on the information paradigm in quantum and gravitational physics. 
{\em J. Phys.: Conf. Ser.} {\bf 880} (2017) 012014.  

\bibitem{Hoehn2} P.A. Hoehn, Quantum theory from rules on information acquisition. 
{\em Entropy} {\bf 19} (2017) 98.

\bibitem{Hoehn3} P.A. Hoehn and C. Wever, Quantum theory from questions. 
{\em Phys. Rev. A} {\bf 95} (2017) 012102.  

\bibitem{Vervoort1} L. Vervoort, Bell's Theorem: Two Neglected Solutions. {\em Found. of Phys.} {\bf 43} (2013) 769. 

\bibitem{Vervoort2} L. Vervoort, Probability theory as a physical theory points to superdeterminism. {\em Entropy} {\bf 21}(9) (2019) 848. 

%

\bibitem{Shannon} C.E. Shannon, Communication in the presence of noise.  
{\em Proc. IRE} {\bf 37} (1949) 10. 

\bibitem{Jerri} A.J. Jerri, The Shannon Sampling Theorem --- Its Various  Extensions and 
Applications: A Tutorial Review.  
{\em Proc. IEEE} {\bf 65} (1977) 1565.   

\bibitem{TDLee} T.D. Lee, Can time be a discrete dynamical variable? 
{\em Phys. Lett.} {\bf 122B} (1983) 217; {\it do.},    
Difference equations and conservation laws.  
{\em J. Statist. Phys.} {\bf 46} (1987) 843. 

%

\bibitem{ElzeRelativPart} H.-T. Elze, Emergent discrete time and quantization: relativistic particle with extradimensions. 
{\em Phys. Lett. A} {\bf 310} (2003) 110. 

%

\bibitem{Visser} A. Van-Brunt and M. Visser, Explicit Baker-Campbell-Hausdorff formulae for some specific Lie algebras. 
{\em Mathematics} {\bf 6} (2018) 135. 

\bibitem{Matone} M. Matone, Closed form of the Baker-Campbell-Hausdorff formula for the generators of semisimple complex Lie algebras. {\em Eur. Phys. J. C} {\bf 76}  (2016) no.11, 610. 

%

\bibitem{tHooftScard} F. Scardigli, G. 't\,Hooft, E. Severino, P. Coda, 
Determinism and Free Will - New Insights from Physics, Philosophy, and Theology. (Springer Nature Switzerland AG, 2019) p.\,28. 

%

\bibitem{AtmanspacherPrimas} H. Atmanspacher and H. Primas, Epistemic and ontic quantum realities. In: A. Khrennikov (ed.), Foundations of Probability and Physics-3. {\em Conf. Proc. Series} {\bf 750} (AIP, Melville, NY, 2005) 49. 

\bibitem{Khrennikov2} A. Khrennikov, I. Basieva, E.M. Pothos and I. Yamato, Quantum probability in decision making from quantum information representation of neuronal states. {\em Scientific Reports} {\bf 8} (2018) 16225. 

\bibitem{Khrennikov1} A. Khrennikov and E. Yurova, Automaton model of protein: dynamics of conformational and functional states. {\em Progress in Biophysics and Molecular Biology} {\bf 130}, Part A (2017) 2. 

\bibitem{Elzehybrid} H.-T. Elze, Linear dynamics of quantum-classical hybrids. \emph{Phys. Rev. A} {\bf 85} (2012) 052109. 

\end{thebibliography}
\end{document}